# Growth of membranes formed by associating polymers at interfaces

*Elena N. Govorun,*[*,†] *Julien Dupré de Baubigny,*[††] *Patrick Perrin,*[††] *Mathilde Reyssat,*[†] *Nadège Pantoustier,*[††] *Thomas Salez,*[†††] *and Cécile Monteux*[*,††]

[†]UMR CNRS Gulliver 7083, ESPCI Paris, PSL Research University, 75005 Paris, France

[††]Sciences et Ingénierie de La Matière Molle, UMR 7615, ESPCI Paris, PSL Research University, CNRS, Sorbonne Universités, 75005 Paris, France

[†††]Université Bordeaux, CNRS, LOMA, UMR 5798, 33405 Talence, France

for Table of Contents use only

# Growth of membranes formed by associating polymers at interfaces

Elena N. Govorun, Julien Dupré de Baubigny, Patrick Perrin, Mathilde Reyssat, Nadège Pantoustier, Thomas Salez, and Cécile Monteux

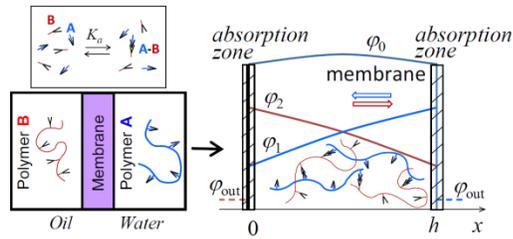

ABSTRACT: Polymer association at liquid-liquid interfaces is a promising way to spontaneously obtain soft self-healing membranes. In the case of reversible bonding between two polymers, the macromolecules are mobile everywhere within the membrane and they can be absorbed into it at both boundaries due to binding to macromolecules of the other type. In this work, we develop the theoretical model of membrane growth based on these assumptions. The asymptotic dependence of membrane thickness on time as $h \sim t^{1/2}$, as typically observed in experiments in a stationary regime, reveals an interdiffusion-controlled process, where the polymer fluxes sustain the polymer absorption at the membrane boundaries. The membrane growth rate is mainly determined by the difference in equilibrium compositions at the boundaries, the association constant, the polymer lengths and mobilities. This model is further used to describe the growth of hydrogel membranes formed via H-bonding of polymers at the interface between a solution of poly(propylene oxide) (PPO) in isopropyl myristate and an aqueous solution of poly(methacrylic acid) (PMAA). The film thickness is measured by reflectometric methods. The dependence of thickness on time can be approximated by the power law $t^{\beta}$, where $\beta \approx 1/2$ for the PMAA solution at pH=3 and decreases with increasing pH and, hence, ionization degree. The growth rate slows down about 25 times for 500-nm-thick films at pH = 5.1 compared to the case of pH = 3. The ionization degree of PMAA solutions was studied by potentiometric methods. Even a small change in ionization was found to influence the growth rate of the film. A slowdown of the film growth for the ionized polymer can be explained by a drop in the composition gradient in the membrane, as is predicted by the proposed model.

1. Introduction

Polymer membranes formed at liquid-liquid interfaces have been actively studied over the last decades[1-20] and they find more and more applications in manufacturing, gas separation, water and air purification, medicine and other areas. The fabrication of membranes often requires the formation of a covalent polymer network. Another type of membranes with self-healing and reconfigurable properties can be obtained by assembling polymer molecules at interfaces.[6-16] Such membranes are hydrogels of associating polymers that are actively studied, in particular, because of their wide range of mechanical properties.[21-24] The hydrogel membranes are also promising in microcapsule production using microfluidic devices.[7,9,11-14]

For membrane formation via interfacial complexation, the interacting species are first dissolved in two separate liquids, such as oil and water or two partially miscible phases. When two solutions are brought into contact, the components diffuse spontaneously toward the interface and they self-assemble through noncovalent interactions. Several experimental studies have reported a stable membrane growth following a $t^{1/2}$ dynamics over time $t$, which suggests that the assembly process is controlled by diffusion.[9-11,16]

Models of growth of interfacial films were first developed for the polycondensation reaction, where the reaction zone is located at the interface between the polymer film and the organic phase and the monomer diffusion through the film controls the process.[25] Several theoretical approaches were used for different types of interfacial polymerization under steady and non-steady conditions.[26-29] Furthermore, the dependence of monomer mobility on film density,[17] the membrane porosity[18] the reversible reaction,[19] and the effect of cross-linking relative to a linear process[21] were analyzed.

In contrast, there are very few studies devoted to the description of interfacial membranes involving noncovalent interactions. In a previous article, we have studied experimentally and theoretically the growth of interfacial membranes that assemble by hydrogen bonds.[10] Using interferometry, we investigated the growth of poly(propylene oxide) (PPO) and poly(methacrylic acid) (PMAA) membranes at the interface between two immiscible phases, water and isopropyl myristate, which contained respectively the H-donor PMAA at pH=3 and the H-bond acceptor PPO. The polymer concentrations in solution remained almost unchanged during the process. We have experimentally found that the growth process is controlled by the molecular mass and the concentration of the PPO chains rather than the ones of the PMAA chains. From the experimental data, we extracted the diffusion coefficient of PPO chains in the membrane and showed that it decreases as the molecular mass decreases and concentration increases. We proposed a model based on the existence of an entropic barrier which slows down the diffusion of free PPO chains through the porous membrane. The pore size was estimated to be smaller than the gyration radius of PPO chain and to decrease with the PPO concentration and molecular mass.

In the present work, we investigate theoretically and experimentally the influence of pH and hence polymer ionization on membrane growth. Earlier, the influence of pH on the membrane properties was described for microcapsules formed via interfacial complexation.[11,12] For the PPO-PMAA microcapsules, a decrease in shear modulus and a subsequent dissolution were observed with increasing pH. For microcapsule shells consisting of two weak polyelectrolytes, a decrease in stiffness by two orders of magnitude was found, whereas the capsule size changed slightly.[12] In this study, we determine the dependence of the PMAA ionization degree on pH in aqueous solutions and measure the film thickness at the interface depending on time. Even weak ionization of the polymer can significantly slow down membrane growth. Since the polymer ionization hinders the associative bonding and produces counterions, this should lead to the formation of more compliant membranes. The previous description of the membrane as a porous solid, where the diffusion of free polymer chains should be faster in membranes with larger pores, does not allow to capture the slowdown of the process. We hence suggest here a new model of membrane growth assuming that all polymer chains are mobile in the considered case of reversible bonding and they can be absorbed into the hydrogel membrane at both boundaries. The whole system is not in thermodynamic equilibrium and we assume that only boundary layers of the membrane can be close to equilibrium with the external solutions. We note that the hydrogel thermodynamics can be analyzed considering a model mixture of hetero-associative polymers in a common solvent, as was recently proposed.[30]

The equilibrium parameters of the hydrogel boundaries are calculated by taking into account reversible bonding between the polymers of different types, the polymer-solvent interactions and the osmotic pressure of counterions. We analyze the influence of polymer ionization on the membrane composition and growth rate and compare the results with our experimental data. The fact that even a small change in ionization degree can greatly slow down the process can be understood in terms of the present thermodynamic approach.

## 2. Theoretical model

We consider a membrane formed via reversible noncovalent bonding (hydrogen bonding) between two polymers that are initially in different solutions. When these two solutions are brought into contact, a membrane starts to grow spontaneously at their joint interface. The hydrogel membrane continues to grow even for very long times, and the membrane thickness $h$ increases with time $t$ approximately as $h \sim t^{1/2}$ if the polymers are almost not ionized.[10]

Such a dependence may be due to the diffusion control of the process with a stationary distribution of reagents in the membrane. Models of the reactive growth of films were well developed for the interfacial polycondensation, where the reaction zone is located at the interface between the polymer film and the organic phase, and the monomer diffusion through the film controls the process in a steady regime.[25-29] However, this model is not directly applicable to the hydrogel membrane, because i. the reversible bonding (reaction) occurs everywhere in the membrane, ii. both polymers are mobile reagents and they can be incorporated into the membrane at the interface from both solutions.

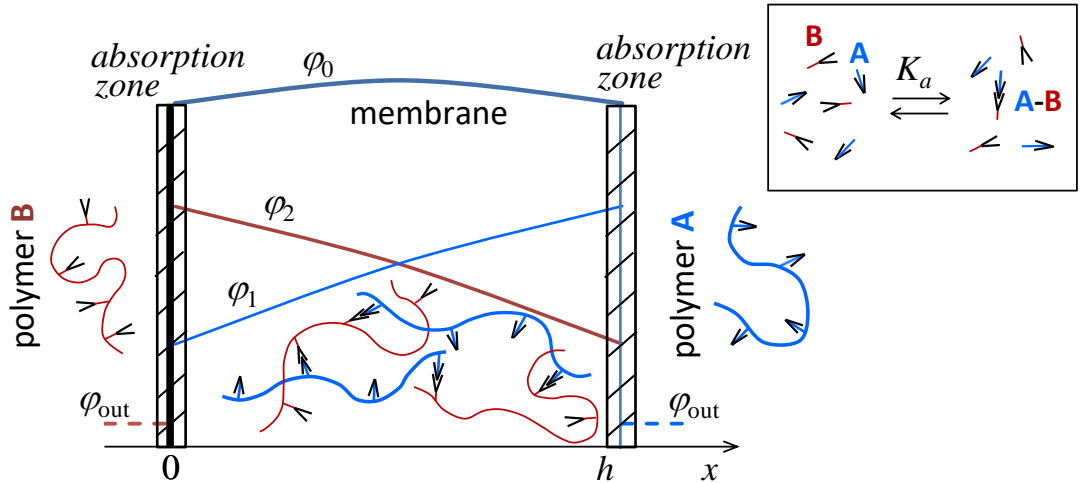

Figure 1. Schematic of a flat membrane between the planes $x = 0$ and $x = h$ separating solutions of two different polymers. The polymer volume fraction in the external solutions, $\varphi_{\text{out}}$, the volume fractions of the A and B polymers, $\varphi_1$ and $\varphi_2$, and the total polymer volume fraction $\varphi_0$ in the membrane are qualitatively shown as functions of the depth $x$ in the membrane. The polymers are absorbed into the gel membrane at both its sides due to associative bonding between the monomer units of A and B types. $K_a$ is the association constant.

We therefore suggest a novel model of the membrane growth. The membrane scheme is presented in Fig. 1. The film is formed at the interface between the solution of the A polymer (right) and the solution of the B polymer (left). The monomer units of type A are able to reversibly associate with the monomer units of type B, their bonding being characterized by the association constant $K_a$. Considering PMMA as the A polymer and PPO as the B polymer, it is worth noting that more types of hydrogen bonding occur in the PPO-PMMA membranes. The hydrogen atoms of the hydroxyl groups of the MMA monomer can bind with the oxygen atoms not only of the PO monomer (an A-B bond) but also of the other hydroxyl groups of MMA (an A-A bond), so that the A polymer chains can be mutually bonded. However, the A-A bonding is assumed to be less important in membrane formation since it does not even interfere with good solubility of pure PMMA in water. Thus, only the A-B bonding is taken into account in the model.

The polymer chains are absorbed into the membrane at the interfaces due to bonding with the polymer chains of the other type. Then, the interdiffusion of polymers through the membrane is necessary to sustain the membrane growth. So, the process can be characterized as interdiffusion controlled.

*2.1 Mathematical description*

The mathematical model of the process can be formulated in a simple form in the particular case of symmetric conditions considering only the free energy of bonding (and not polymer-solvent interactions) as the interaction energy. The total monomer unit concentration $c_0$ in the membrane is assumed to be constant and the interdiffusion is described in the same way as for polymer melts. We also assume that only small fractions of monomer units of both types are associated and that the self-diffusion coefficient $D$ of polymer chains in the gel membrane is approximately the same as in a solution of the same density but without association. The latter condition should be valid if the lifetime of reversible bonds is much smaller that the relaxation time of the polymer chains.

The membrane is formed by the neutral A and B polymers of length $N$ with the monomer units of volume $\upsilon$. The monomer unit concentrations of the A polymer in the right solution and B polymer in the left solution are equal to $c_{\text{out}}$. The monomer unit concentrations of the A and B polymers in the membrane are denoted by $c_1$ and $c_2$, respectively. It is assumed that the membrane is flat and that these concentrations depend only on the coordinate $x$ (Fig. 1) as: $c_1 = c_1(x)$, $c_2 = c_2(x)$. The A and B polymer volume fractions, $\varphi_1$ and $\varphi_2$, are related to the corresponding concentrations as $\varphi_1 = c_1 \upsilon$, $\varphi_2 = c_2 \upsilon$, and $\varphi_{\text{out}} = c_{\text{out}} \upsilon$. The total polymer volume fraction $\varphi_0 = \varphi_1 + \varphi_2$ is a constant, the solvent volume fraction is $\varphi_s = 1 - \varphi_0$.

The polymer hydrogel is considered as a polymer mixture with the free energy consisting of the mixing entropy contribution and the interaction energy. We assume that the association energy dominates among all the interaction types and the interaction energy density is given by the contribution $f_{\text{h.b.}}$ of hydrogen bonding only. The bonding free energy includes the energy of bonds and the entropic terms. For the equilibrium state with respect to the number of bonds, the bond concentration $c_{\text{h.b.}}$ is related to the concentrations of non-bonded monomer units, $c_{1(\text{rest})}$ and $c_{2(\text{rest})}$, as $c_{\text{h.b.}} = K_a c_{1(\text{rest})} c_{2(\text{rest})}$.[30,31] The association constant is related to the bond energy $\varepsilon$ (in units of $k_B T$) and to the effective bond volume $\upsilon_b$ as $K_a = \upsilon_b e^{|\varepsilon|}$. If the fraction of bonded monomer units is small, then $c_{1(\text{rest})} \approx c_1 = \varphi_1/\upsilon$, $c_{2(\text{rest})} \approx c_2 = \varphi_2/\upsilon$, and the free energy density is approximately equal to

$$\frac{f_{\text{h.b.}} \upsilon}{k_B T} \approx -c_{\text{h.b.}} \upsilon = -\widetilde{K}_a \varphi_1 \varphi_2, \tag{1}$$

where $\widetilde{K}_a = K_a/\upsilon$.

We describe the interdiffusion of polymers in the hydrogel by the same equations as for the binary melt,[32-34] where the interaction energy density is given by the expression $f_{\text{int}} \upsilon/(k_B T) = \chi \widetilde{\varphi}(1 - \widetilde{\varphi})$, $\chi$ is the Flory-Huggins parameter of interactions between the A and B monomer units, and $\widetilde{\varphi}$ is the volume fraction of the A polymer. The contribution of interactions into the chemical potential of the first polymer is $\mu_{\text{int}} = \upsilon \partial f_{\text{int}}/\partial \widetilde{\varphi} = \chi(1 - 2\widetilde{\varphi})$.

In the hydrogel, the fraction of the A polymer is $\rho = \varphi_1/\varphi_0$ and the bonding free energy can be expressed as $f_{\text{h.b.}} \upsilon/(k_B T) = -\widetilde{K}_a \varphi_0^2 \rho(1 - \rho)$, then the chemical potential of the A polymer is equal to $\mu_{\text{h.b.}} = \upsilon \partial f_{\text{h.b.}}/\partial \varphi_1 = -\widetilde{K}_a \varphi_0 (1 - 2\rho)$. The dependences of the chemical potential on the composition of the melt and hydrogel become equivalent if $\chi$ is equal to $-\widetilde{K}_a \varphi_0$. If the polymer interdiffusion in the hydrogel can be described as in the melt, the diffusion equation takes the form

$$\frac{\partial \rho}{\partial t} = D \frac{\partial}{\partial x}\left(\left(1 + 2\widetilde{K}_a N \varphi_0 \rho(1-\rho)\right)\frac{\partial \rho}{\partial x}\right) \tag{2}$$

and the flux of the A polymer fraction can be represented as

$$J_\rho = -D\left(1 + E_0 \rho(1-\rho)\right)\frac{\partial \rho}{\partial x}, \tag{3}$$

where $E_0 = 2\widetilde{K}_a N \varphi_0$. It is worth noting that the interdiffusion rate is underestimated in the present consideration because the model assumptions correspond, in fact, to an immobile solvent.

The absorption of polymer chains into the membrane at the boundaries proceeds mainly due to the association with the polymer of the other type. The flux of the B polymer at the left boundary and the flux of the A polymer at the right boundary should be proportional to the difference in the chemical potentials in the gel and outside. Let $\rho^*$ be the fraction of the A polymer at the left interface, assumed to be in equilibrium with the external solution. The equilibrium conditions imply the equalities of osmotic pressures and chemical potentials which determine the value of $\rho^*$ depending on the polymer volume fraction $\varphi_{\text{out}}$ in the external solution. The same value $\rho^*$ is reached by the equilibrium fraction of the B polymer at the right interface. Taking into account only the bonding energy contribution in the chemical potential, we can write the boundary conditions for the flux of the A polymer fraction in the form

$$-J_\rho|_{x=0} = u(\rho - \rho^*)|_{x=0}, \qquad -J_\rho|_{x=h} = u(1 - \rho^* - \rho)|_{x=h}, \tag{4}$$

where the coefficient $u$ characterizes the absorption rate. It should be proportional to the association constant and to the polymer concentration in the external solution: $u \sim \widetilde{K}_a \varphi_{\text{out}}$.

If the membrane grows slowly and a stationary dependence of the gel composition on the coordinate is established, the flux $J_\rho$ does not depend on the coordinate and we can write

$$-J_\rho = D\left(1 + E_0 \rho(1-\rho)\right)\frac{\partial \rho}{\partial x} = C, \tag{5}$$

where the parameter $C$ is determined only by the thickness $h$. Introducing the function $\Phi(\rho)$ by the equation

$$(1 + E_0\rho(1-\rho))\frac{\partial \rho}{\partial x} = \frac{\partial \Phi(\rho(x))}{\partial x}, \tag{6}$$

we can find $\Phi(\rho) = \rho + E_0(\rho^2/2 - \rho^3/3)$ and, then, the equation (5) for the function $\rho(x)$ can be rewritten as $\Phi(\rho(x)) - \Phi(\rho_{\text{left}}) = Cx/D$, where $\rho_{left} = \rho(0)$ and the zero coordinate $x = 0$ is taken at the left surface of the membrane. It follows from the boundary conditions (4) for $J_\rho = const$ that $u(\rho_{\text{left}} - \rho^*) = u(1 - \rho^* - \rho_{\text{right}})$, where $\rho_{\text{right}} = \rho(h)$. Then, $\rho_{\text{right}} = 1 - \rho_{\text{left}}$ and Eq. (5) can be written at $x = h$ as the implicit dependence of $\rho_{\text{left}}$ on $C$:

$$\Phi(1 - \rho_{\text{left}}) - \Phi(\rho_{\text{left}}) = \frac{C}{D}h. \tag{7}$$

Equations (5)-(7) also implicitly determine the corresponding dependence $\rho(x)$. From the boundary conditions (5), one has

$$\rho_{\text{left}} = \frac{C}{u} + \rho^* \tag{8}$$

that permits to find $\rho_{\text{left}}$ and $C$ for a given $h$ from the system of equations (7) and (8): $\rho_{\text{left}} = \rho_{\text{left}}(h)$, $C = C(h)$.

The increase in the membrane thickness $h$ comes from the absorption of polymer chains at both surfaces and can be related to the boundary conditions:

$$\frac{dh}{dt} = -J_\rho|_{x=0} - J_\rho|_{x=h} = 2C(h). \tag{9}$$

Then, the membrane thickness $h(t)$ can be found from the equations

$$\int_0^h \frac{dx}{C(x)} = 2t \quad \text{or} \quad \int_0^h \frac{dx}{u(\rho_{\text{left}}(x) - \rho^*)} = 2t. \tag{10}$$

With increasing $h$, the value of $C$ decreases to satisfy Eq. (7) and then the fraction $\rho_{\text{left}}$ of the A polymer at the left boundary tends to $\rho^*$, i.e., $\rho_{left} - \rho^* \ll 1$ (Eq. (8)); the same is valid for the fraction of B polymer at the right boundary. Asymptotically, the equation (7) takes the form

$$\frac{C}{D}h \approx (1-2\rho^*)\left(1+\frac{E_0}{6}f_{\rho^*}\right) = \Delta\rho^*\left(1+\frac{E_0}{6}f_{\rho^*}\right), \tag{11}$$

where $\Delta\rho^* = 1 - 2\rho^*$ is the difference in the equilibrium gel composition at the film boundaries and $f_{\rho^*} = 1 + 2\rho^* - 2(\rho^*)^2$. The membrane thickness $h$ can be found from Eq. (10) and (11):

$$h \approx 2\sqrt{D\Delta\rho^*\left(1+\frac{E_0}{6}f_{\rho^*}\right)t}. \tag{12}$$

Thus, the asymptotic power law $h \sim t^{1/2}$ is predicted revealing a diffusion-controlled process. The growth rate is determined by the difference in the equilibrium gel composition at the film boundaries, the association constant, the gel density, and the mobility and length of macromolecules. Asymptotically (Eq. (12)), the growth rate does not depend on the absorption rate $u$.

For large values of $E_0$, $E_0 = 2\widetilde{K}_a N\varphi_0 \gg 1$, the thickness $h$ (Eq. (12)) can be estimated as

$$h \approx \frac{2}{\sqrt{3}}\sqrt{D\widetilde{K}_a N\varphi_0 t} \quad \text{if} \quad \rho^* \ll 1,\ \Delta\rho^* \approx 1;$$

$$h \approx \sqrt{2D\Delta\rho^*\widetilde{K}_a N\varphi_0 t} \quad \text{if} \quad \rho^* \approx 0.5,\ \Delta\rho^* \ll 1. \tag{13}$$

The second case corresponds to a small difference $\Delta\rho^*$ in equilibrium gel compositions at the boundaries. In that case, the growth rate is very sensitive to the value of $\Delta\rho^*$ that depends on the parameters of external solutions and the molecular characteristics.

*2.2 Hydrogel thermodynamics: effect of polymer ionization*

We consider a process of membrane growth under constant polymer concentrations in the external solutions and analyze the influence of ionization of the A polymer under an external changing pH. It is assumed that the A polymer dissolved in the first solvent does not penetrate into the second solvent, whereas the B polymer dissolved in the second solvent can be absorbed into the membrane immersed in the first one despite the poor solvent conditions for this polymer. Then, the membrane (Fig. 1) should be immersed in the first solvent with the left boundary coinciding with the interface between the two solvents (at $x = 0$).

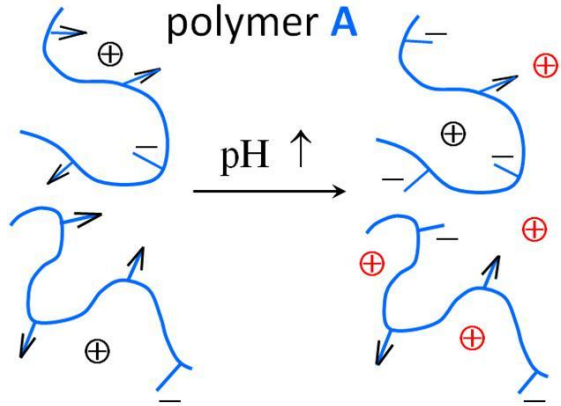

Figure 2. Schematic of the A polymer ionization with increasing pH. The polymer is slightly ionized without the addition of alkali (left). In the presence of dissociated alkali, more monomer units are ionized. Their counterions $H^+$ (black circles) are partially neutralized by the negative ions $OH^-$ of the alkali to form water molecules (not shown). For a small amount of alkali, the concentration of charged monomer units is equal to the total concentration of positive ions.

The value of pH is controlled by the monobasic alkali concentration $c_e$ in the external solution of the A polymer. The concentrations of positive ions of dissociated alkali in the membrane and solution are $c_e^+$ ($c_e^+ = c_e$) and $c_{in}$, respectively. The monomer units of the A polymer dissociate into counterions $H^+$ and charged monomer units that are not capable of hydrogen bonding (Fig. 2). Let $\alpha$ be the ionization degree of monomer units of the A polymer. It is assumed to be the same in the external solution and in the hydrogel film. The film remains electrically neutral due to the presence of mobile ions of opposite sign. For a small amount of the alkali added (no free $OH^-$ ions), the total concentration of mobile positive ions is $c_{in} + c_{H^+} = c_c$, where $c_c = \alpha\varphi_1/\upsilon$ is the concentration of charged monomer units. The same condition: $c_e + c_{H^+(out)} = c_{c(out)}$, $c_{c(out)} = \alpha \varphi_{1out}/\upsilon$, is valid for the external solution.

We consider the membrane as a mixture of two polymers of lengths $N_1$ and $N_2$ in a common solvent. The free energy density $f$ of a homogeneous mixture contains the entropic contribution $f_{mix}$ of the polymers and solvent, the entropic contribution of ions $f_{ion}$, the free energy of hydrogen bonding $f_{h.b.}$, and the enthalpic energy of polymer-solvent interactions $f_{int}$:

$$f = f_{mix} + f_{ion} + f_{h.b.} + f_{int}, \qquad (14)$$

where

$$\frac{f_{mix}\upsilon}{k_B T} = \frac{\varphi_1}{N_1}\ln\frac{\varphi_1}{e} + \frac{\varphi_2}{N_2}\ln\frac{\varphi_2}{e} + (1-\varphi_1-\varphi_2)\ln\frac{1-\varphi_1-\varphi_2}{e}. \qquad (15)$$

The entropic contributions of the mobile positive ions of the alkali and polymer can be written as

$$\frac{f_{ion}\upsilon}{k_B T} = \varphi_{in}\ln\frac{\varphi_{in}}{e} + (\alpha\varphi_1 - \varphi_{in})\ln\frac{\alpha\varphi_1 - \varphi_{in}}{e}, \qquad (16)$$

where $\varphi_{in} = c_{in}\upsilon$. The excluded volume interactions of the ions are neglected in Eq. (16).

The concentration of hydrogen bonds $c_{h.b.}$ depends on the polymer ionization degree, since charged monomer units cannot form bonds. In the equilibrium state, the concentration $c_{h.b.}$ is determined by the concentration of non-bonded monomer units as $c_{h.b.} = K_a c_{1(rest)} c_{2(rest)}$. If the fraction of bonded monomer units is small, then $c_{1(rest)} \approx (1-\alpha)c_1$, $c_{2(rest)} \approx c_2$ and the free energy of bonding

$$\frac{f_{h.b.}\upsilon}{k_B T} \approx -c_{h.b.}\upsilon = -\widetilde{K}_a(1-\alpha)\varphi_1\varphi_2. \qquad (17)$$

The solvent is assumed to be good for the A polymer (with the Flory-Huggins parameter $\chi_{As} = 0$) and poor for the B polymer (with the Flory-Huggins interaction parameter $\chi_{Bs} = \chi_2 > 0.5$). The B polymer-solvent interactions are described by the expression

$$f_{int}\upsilon = k_B T \chi_2 \varphi_2 \varphi_s. \qquad (18)$$

The free energy is written similarly to the one analyzed for polyelectrolyte solutions and gels, except the bonding term. We do not consider the elastic free energy of polymer chains in the total free energy because the bonds between them are temporary and conformational deformations can relax on experimental time scales (minutes and hours). The electrostatic interaction energy of charged monomer units is neglected for the considered case of small polymer ionization.

The chemical potentials of the A and B polymers (per monomer unit) are

$$\mu_1 = v\frac{\partial f}{\partial \varphi_1}, \quad \mu_2 = v\frac{\partial f}{\partial \varphi_2}. \tag{19}$$

We assume that the growing membrane is locally homogeneous, that is the hydrogel is stable with respect to concentration fluctuations. Using the notation $\varphi_3=\varphi_{in}$, the free energy (14) can be written as a function of three independent variables $\varphi_1$, $\varphi_2$, $\varphi_3$, and the stability conditions can be found by analyzing the matrix $\hat{m}$ of the second derivatives of the free energy density, $m_{ij} = \frac{\partial^2 f}{\partial \varphi_i \partial \varphi_j}$, $i,j = 1, 2, 3$. The condition for the determinant to be positive should be satisfied:

$$d_3 = m_{33}d_2 - m_{13}^2 m_{22} > 0, \quad \text{where } d_2 = m_{11}m_{22} - m_{12}m_{21}. \tag{20}$$

The matrix elements $m_{ij}$ are

$$m_{11} = \frac{1}{N_1 \varphi_1} + \frac{1}{\varphi_s} + \frac{\alpha^2}{\alpha\varphi_1 - \varphi_{in}}, \quad m_{22} = \frac{1}{N_2 \varphi_2} + \frac{1}{\varphi_s} - 2\chi_2$$

$$m_{33} = \frac{1}{\alpha\varphi_1 - \varphi_{in}} + \frac{1}{\varphi_{in}}, \quad m_{12} = m_{21} = \frac{1}{\varphi_s} - \widetilde{K}_0(1-\alpha) - \chi_2 \tag{21}$$

$$m_{13} = m_{31} = -\frac{\alpha}{\alpha\varphi_1 - \varphi_{in}}, \quad m_{23} = m_{32} = 0$$

Eq. (20) determines the possible values of the volume fractions $\varphi_1$, $\varphi_2$, $\varphi_{in}$ in a stable homogeneous hydrogel.

As long as the membrane continues to grow, the whole system is not in thermodynamic equilibrium. We consider a stationary regime of membrane growth in which the interdiffusion is assumed to be slow and the parameters of the boundary layers established almost in the thermodynamic equilibrium with the external solutions. It is assumed that there is no admixture of the A polymer is the second solution and the B polymer in the first one and that the solvents are good for their polymers, i.e., the Flory-Huggins parameters of the polymer-solvent interactions are equal to zero.

The free energies of the first solution with the polymer volume fraction $\varphi_{1out}$ and second solution with the polymer volume fraction $\varphi_{2out}$ include only the entropic terms:

$$\frac{f_{1out}\upsilon}{k_BT} = \frac{\varphi_{1out}}{N_1}\ln\frac{\varphi_{1out}}{e} + (1-\varphi_{1out})\ln\frac{1-\varphi_{1out}}{e} + \varphi_{H^+(out)}\ln\frac{\varphi_{H^+(out)}}{e} + \varphi_e\ln\frac{\varphi_e}{e}, \quad (22)$$

$$\frac{f_{2out}\upsilon}{k_BT} = \frac{\varphi_{2out}}{N_2}\ln\frac{\varphi_{2out}}{e} + (1-\varphi_{2out})\ln\frac{1-\varphi_{2out}}{e}, \quad (23)$$

where $\varphi_{H^+(out)} = \alpha\varphi_{1out} - \varphi_e$, $\varphi_e = c_e\upsilon$. The chemical potentials of the polymers in their solutions and the osmotic pressures are:

$$\mu_{1out} = \upsilon\frac{\partial f_{1out}}{\partial \varphi_{1out}}, \quad p_{1out} = f_{1out} - \varphi_{1out}\frac{\partial f_{1out}}{\partial \varphi_{1out}} - \varphi_e\frac{\partial f_{1out}}{\partial \varphi_e}, \quad (24)$$

$$\mu_{2out} = \upsilon\frac{\partial f_{2out}}{\partial \varphi_{2out}}, \quad p_{2out} = f_{2out} - \varphi_{2out}\frac{\partial f_{2out}}{\partial \varphi_{2out}}.$$

The osmotic pressure in the hydrogel is

$$p = f - \varphi_1\frac{\partial f}{\partial \varphi_1} - \varphi_2\frac{\partial f}{\partial \varphi_2} - \varphi_{in}\frac{\partial f}{\partial \varphi_{in}}. \quad (25)$$

The conditions of thermodynamic equilibrium with respect to water swelling of the gel and to A polymer absorption at the right boundary are

$$p = p_{1out} \quad (26)$$

$$\mu_1 = \mu_{1out} \quad (27)$$

The condition (26) takes into account the osmotic pressures of mobile ions inside and outside the film, where their difference should be equilibrated by the other pressure contributions.[35]

The conditions of the thermodynamic equilibrium with respect to the ion penetration into the gel is[35]

$$\frac{\partial f}{\partial \varphi_{in}} = \frac{\partial f_{1out}}{\partial \varphi_e}, \tag{28}$$

that gives the Donnan-type equation for the ion concentrations:

$$\frac{\varphi_{in}}{\alpha\varphi_1 - \varphi_{in}} = \frac{\varphi_e}{\alpha\varphi_{1out} - \varphi_e}, \tag{29}$$

leading to $\varphi_{in} = \varphi_e(\varphi_1/\varphi_{1out})$.

Below, we calculate the possible equilibrium parameters of the gel at the right and left boundaries for the neutral and weakly charged A polymer to reveal the influence of polymer ionization on the gel composition. If the polymers are neutral ($\alpha = 0$), they do not impose any limitation on the positions of mobile ions due to electric neutrality, and the ion contribution to the free energy (14) can be omitted, that is $f_{ion} = 0$. If the A polymer is charged, the entropic contributions of ions appear both in the osmotic pressure and in the chemical potential of the A polymer.

In the case we are interested in, when the growth rate slows down, concentrations of ions can be estimated from experimental parameters. Such a behavior is observed at pH values between 4 and 5 corresponding to the concentration of hydrogen ions, which is many times less than the concentration of dissociated monomer units. This allows us to set $\varphi_{H^+} = \varphi_{H^+(out)} = 0$. Then, $\varphi_{in} = \alpha\varphi_1$, $\varphi_e = \alpha\varphi_{1(out)}$, and the entropic contribution of ions takes the form $f_{ion}\upsilon = k_B T \alpha\varphi_1 \ln(\alpha\varphi_1/e)$.

In Fig. 3, the possible states of the hydrogel satisfying the condition of gel equilibrium with respect to water swelling (Eq. (26)) are presented for the neutral polymers ($\alpha = 0$) and charged A polymer ($\alpha = 0.03$). A homogeneous mixture is unstable (Eq. (20)) for very small or large values of the fraction $\rho = \varphi_1/\varphi_0$, and only the region of stability is shown. The chemical potential $\mu_1$ of the A polymer increases with $\rho$, whereas the chemical potential in the outer solution is described by a horizontal line (Fig. 3a). The equilibrium value of the gel composition at the right boundary $\rho_{right}$ corresponds to a point of intersection (Eq. (27)), where the contributions of polymer mixing, hydrogen bonding, polymer-solvent interactions, and mobile ions (for $\alpha > 0$) are equilibrated.

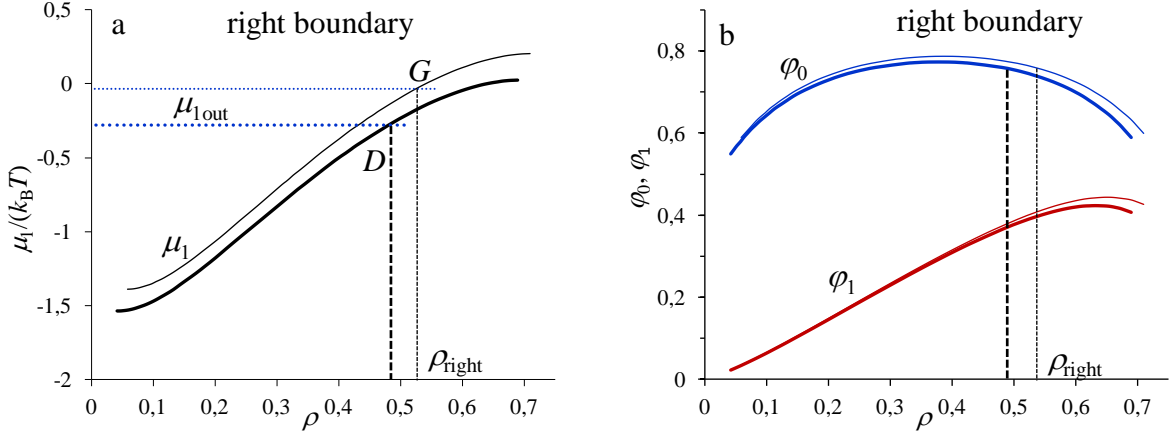

Figure 3. (a) The chemical potential $\mu_1$ of the A polymer and (b) the total polymer volume fraction $\varphi_0$ and the volume fraction of the A polymer $\varphi_1$ as functions of the gel composition $\rho = \varphi_1/\varphi_0$ under the osmotic pressure $p = p_{1out}$ for the neutral polymers ($\alpha = 0$, thin curves) and for the ionized A polymer ($\alpha = 0.03$, thick curves). The dotted horizontal lines (Fig. 3a) describe the chemical potential $\mu_{1out}$ of the A polymer in the external solution, their cross points with the curves, $D$ and $G$, give the equilibrium compositions $\rho_{right}$ at the right boundary. Other parameters: $\widetilde{K}_a = 3.3$, $N_1 = 100$, $N_2 = 70$, $\chi_2 = 0.75$, $\varphi_{1out} = 0.01$.

The dependences of the total polymer volume fraction $\varphi_0 = \varphi_1 + \varphi_2$ and the volume fraction of the A polymer $\varphi_1$ on $\rho$ under the equal osmotic pressures, $p = p_{1out}$, are presented in Fig. 3b. The equilibrium values of the volume fractions are given by the intersection of the curves with the vertical line at $\rho = \rho_{right}$. Ionization of the A polymer leads to a decrease in the equilibrium fraction $\rho_{\text{right}}$ while the total polymer volume fraction $\varphi_0$ is almost unchanged. A larger fraction of the B polymer in the gel with $\alpha > 0$ is necessary to compensate the entropic contribution of mobile ions by hydrogen bonding.

The left boundary of the membrane coincides with the interface between two immiscible solvents and the boundary position can only be shifted by penetration of the B polymer chains into the membrane situated in the first solvent. By minimizing the sum of the free energies of the boundary layer and the second solution with respect to the boundary position, one can obtain

$$\omega = \mu_2 + pv = \mu_{2\text{out}} + p_{2\text{out}} v = const \tag{30}$$

Possible states of the gel that satisfy Eq. (30) and the stability condition (21) are presented in the Fig. 4 as the dependences of the total volume fraction $\varphi_0$ and the volume fraction of the B polymer $\varphi_2$ on $\rho$. The range of possible gel compositions is quite narrow. A higher total polymer concentration corresponds to a higher concentration of the B polymer. Since interdiffusion of the polymers in the membrane is assumed to be slow, the B polymer can be accumulated at the left interface until $\varphi_2$ reaches a maximum possible value corresponding to the leftmost point of the curve. The ionization degree of a few percent leads to very small increase in the equilibrium composition $\rho_{\text{left}}$ in comparison with the change in the equilibrium composition $\rho_{\text{right}}$ at the right boundary. Therefore, a change in the equilibrium gel compositions at the boundaries $\Delta\rho^* = \rho_{\text{right}} - \rho_{\text{left}}$ with ionization is mainly controlled by the value $\rho_{\text{right}}$ at the right boundary.

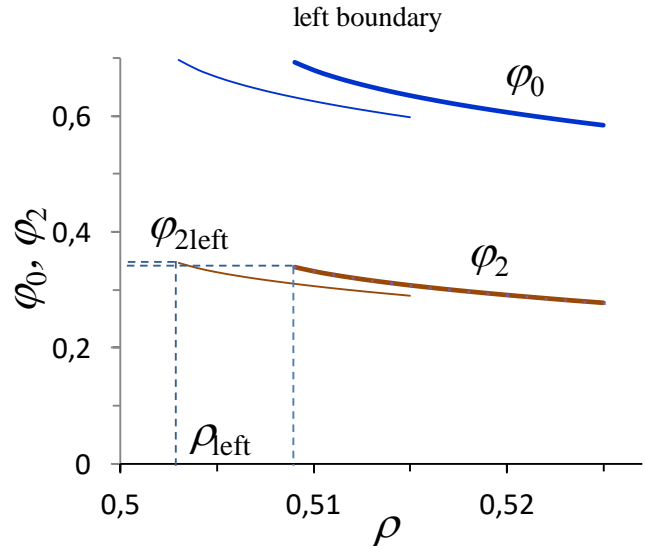

Figure 4. The dependence of the total polymer volume fraction $\varphi_0$ (blue curves) and the volume fraction of B monomer units $\varphi_2$ (brown curves) satisfying Eqs. (20) and (30) on composition $\rho = \varphi_1/\varphi_0$ for the ionization degree $\alpha = 0$ (thin curves) and 0.03 (thick curves). Other parameters: $\widetilde{K}_a = 3.3$, $N_1 = 100$, $N_2 = 70$, $\chi_2 = 0.75$, $\varphi_{2\text{out}} = 0.1$.

## 3. Experimental study of the thickness of PMAA/PPO membranes as a function of ionization degree

### 3.1 Materials and methods

We use PMAA as an H-bond donor of molar mass 100000 K from Polysciences and PPO as an H-bond acceptor, of molar mass 4000, purchaised from Sigma Aldrich and used as received. PMAA 1% is dissolved in water and the pH of the aqueous solutions is adjusted between 3 and 5 using aliquots of NaOH solutions at 0.1 M using a pH meter (pH M 250 ion analyzer, Meterlab, Radiometer Copenhagen). PPO is dissolved in IPM (isopropyl myristate). Molar masses and polydispersity indexes (PDI) were measured on a size exclusion chromatograph (SEC, Viscotek GPC max VE 2001; TDA 302 triple detector array, system with triple detection using 0.2 M $NaNO_3$ aqueous solution or THF containing 2% trimethylamine for PMAA and PPO samples, respectively).

Using SEC in aqueous solution, we measure for PMAA 100K a number-averaged molar mass $M_n$ = 100 000 g/mol and PDI around 2. Using SEC in THF we find that PPO 4000 has a number-averaged molar mass $M_n$ = 3000 g/mol and PDI = 1.97.

To obtain our interfacial membranes, the PMAA and PPO solutions are put in contact in a beaker. The thickness of the membrane is obtained by two methods. The first method consists in measuring the membrane thickness in situ using a reflected light microscope mounted with an optical spectrometer (specim V8 connected to a camera). We focus white light on the oil-water interface. The spectrometer provides the reflected intensity as a function of wave length from which we deduce the membrane thickness. The second method consists of removing the membrane from the liquid, leaving it on a glass slide, and measuring its thickness ex situ, with an optical interferometric profilometer (Microsurf 3D Fogale Nanotech).

The potentiometric measurements were performed using a digital pH meter (pH M 250 ion analyzer, Meterlab, Radiometer Copenhagen) using a combined glass/calomel electrode. The electrode system was daily calibrated using three buffers (pH 4, 7 and 10). The potentiometric titrations were carried out with sodium hydroxide (NaOH) solution (0.2M), according to the PMAA concentration (0.1 mol/L). The measurements were made at 25°C, with constant stirring. The pH readings were carried out after sufficient stabilization periods. The dependence of the PMAA ionization degree $\alpha$ on pH was determined by using the potentiometric titration data as:

$$\alpha = \frac{c_{\text{NaOH}} + c_{\text{H}^+}}{c_{\text{MAA}}}, \tag{31}$$

where $c_{\text{NaOH}}$ and $c_{\text{H}^+}$ are the molarities of added NaOH and free hydrogen ions, respectively, and $c_{\text{MAA}}$ is the molarity of monomer units (as each monomer unit bears one carboxylic group). The ionization degree defined above takes into account the fact that the molarity of hydroxide ions is very low for the cases of weak and moderate polymer ionization. With this definition, one has $\alpha = 1$ at complete ionization.

**3.2 Experimental results and discussion**

The influence of pH values on the hydrogel properties may be due to the polymer ionization with the concomitant localization of counter ions. The ionization degree $\alpha$ of the PMAA solution (1 wt %) depending on pH (Fig. 5) was determined from the data on the NaOH titration (Eq. (31)). In pure water, the polymer solution possesses pH ≈ 2.8 that corresponds to the fraction $\alpha \approx 0.01$ of ionized monomer units. With increasing pH, the ionization reaches approximately 10% at pH = 5.

The detailed study of the PMAA ionization depending on pH and ionic strength was previously done with an accent on possible conformational transitions of PMAA chains.[36] The polymer is expected to be cross-linked by hydrogen bonds at weak ionization and to have loose conformations at high ionization. Such a transition can be detected, in particular, by a change in the linear dependence of pH on $\log(1-\alpha)/\alpha$. In our observations, this change is not pronounced (see inset in Fig. 5).

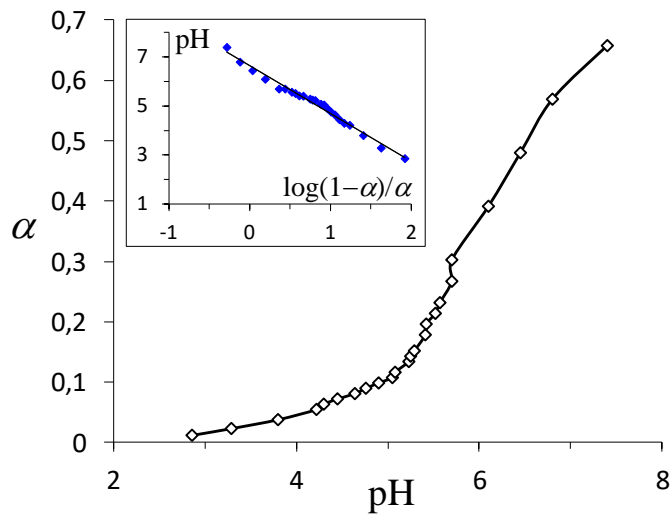

Figure 5. Ionization degree of PMAA (markers) vs pH determined from the titration data (Eq. (31)) for the NaOH solution added to an aqueous solution of PMMA at 25°C, the solid curve is a guide to the eye. This dependence is close to linear in the axes $\log(1-\alpha)/\alpha$ and pH (inset) and can be approximated by the formula $pH = -0.665 \log(1-\alpha)/\alpha + 6.65$ (blue line).

The thickness evolution of the PMAA/PPO membranes as a function of time for different pH values is presented in Fig. 6. Similarly to our previous article,[10] we find that the thickness grows as $t^\beta$ with $\beta = \frac{1}{2}$ at pH = 3. The fitting of the exponent $\beta$ demonstrates its decrease with increasing $\alpha$ and pH to approximately $\beta \approx 0.34$ at pH = 5.1 (inset of Fig. 6). As it can be estimated for the films of 500 nm thickness, the growth rate reduces 1.5 times at pH = 4.5 and 25 times at pH = 5.1 compared to the growth rate at pH=3. At higher values of pH, the formation of membranes is not observed.

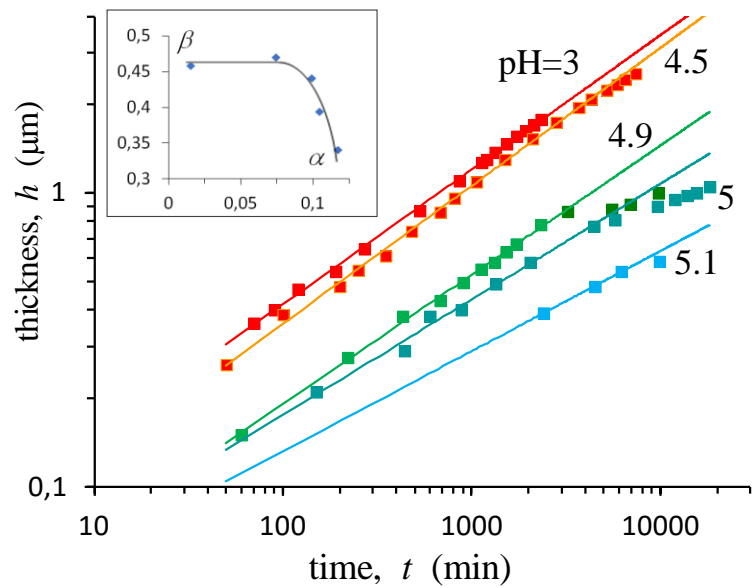

Figure 6. Membrane thickness $h$ vs time $t$ in logarithmic scales for the different values of pH (markers). The concentrations of PPO ($M_n$ = 3000 g/mol) in IPM and PMAA ($M_n$ = 100000 g/mol) in water are 1 wt %. The lines approximate the markers by the power-law dependence $h = \text{const} \cdot t^\beta$. Only the data for $t < 3000$ min are taken into account for pH ≤ 5 and the first three points for pH = 5.1. The inset shows the dependence of the exponent $\beta$ on the ionization degree $\alpha$ (markers). The solid line is a guide to the eye.

We explain the pronounced effect of the pH value on the growth rate by changes in gel composition at the boundaries. In the model, the gradient of gel composition is the driving force of polymer interdiffusion in the membrane. The growth rate at late stages (see Eq. (12)) is also determined by the association constant, the membrane density, and the chain length and mobility. The analysis of the solution predicts that a small change in the difference $\Delta\rho^*$ in equilibrium composition of the gel at the boundaries can significantly affect the growth rate at small values of this difference, i.e., $\Delta\rho^* \ll 1$. In that case, the membrane thickness $h \sim \sqrt{\Delta\rho^*}$ as obtained from Eq. (13).

The theoretical dependences of the membrane thickness $h$ on time are presented in Fig. 7 for the different values of the difference $\Delta\rho^*$ and fixed other parameters. A slight slowdown is observed when the value of $\Delta\rho^*$ decreases from 1 to 0.8 (curves 1 and 2). With further decrease in $\Delta\rho^*$ to small values, the slowdown becomes remarkable (curves 3 and 4) and it is observed both at the initial stage and for large times. The dependence $h \sim t^{1/2}$ is established at approximately the same membrane thickness for the different values of $\Delta\rho^*$. Such a regime corresponds to almost constant values of the gel compositions at the boundaries ($\rho_{\text{left}} - \rho^*$ is close to zero).

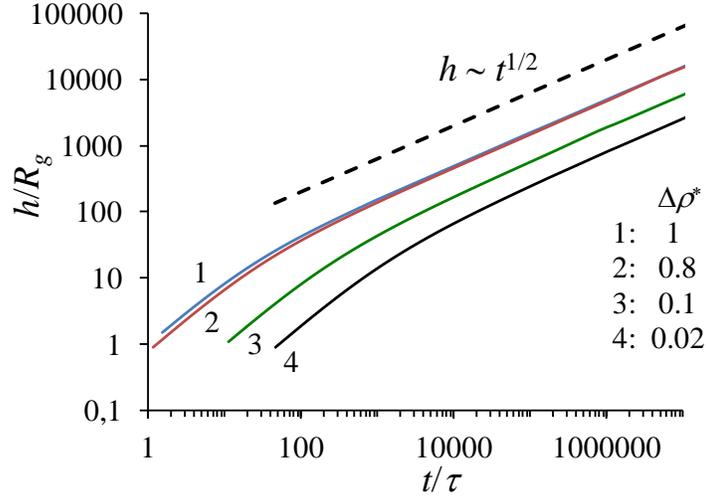

Figure 7. The dependence of the membrane thickness $h/R_g$ on time $t/\tau$ in double logarithmic scale calculated from Eqs. (11)-(13) for the different values of the equilibrium gel composition $\rho^*$ at the boundaries and correspondingly the difference $\Delta\rho^*$: $\Delta\rho^* = 1$ and $\rho^* = 0$ (1), $\Delta\rho^* = 0.8$ and $\rho^* = 0.1$ (2), $\Delta\rho^* = 0.1$ and $\rho^* = 0.45$ (3), $\Delta\rho^* = 0.02$ and $\rho^* = 0.49$ (4). The dashed line indicates the power-law dependence $h \sim t^{1/2}$. The parameters: $E_0 = 2\widetilde{K}_a N\varphi_0 = 30$, $uR_g/D = 1$, and $R_g$ is the gyration radius of polymer chain with $\tau = R_g^2/D$ the characteristic time of chain displacement over the distance $R_g$.

The previous model used for the description of membranes of associating polymers[10] predicted the diffusion-controlled regime of growth $h \sim t^{1/2}$ and explained the dependence of the growth rate on the polymer concentration in solution. However, the idea of the membrane as a porous solid could not explain the slowdown in polymer diffusion under the same concentrations of polymers which become weakly ionized. The new model makes it possible to preserve the diffusion-controlled nature of the process and introduce some control of the process by external parameters through boundary conditions.

The difference in equilibrium gel compositions at the boundaries depending on the ionization degree can be found using the thermodynamic analysis of the hydrogel considered as a polymer mixture in a common solvent. With increasing the ionization degree $\alpha$, the equilibrium fraction $\rho_{\text{right}}$ at the right boundary decreases, whereas the fraction $\rho_{\text{left}}$ at the left boundary slightly increases (Figs. 3 and 4). So, the difference in the equilibrium values of $\rho$ at the right and left boundaries, $\Delta\rho = \rho_{\text{right}} - \rho_{\text{left}}$, is maximum for the non-ionized polymer ($\alpha = 0$) and it decreases with increasing $\alpha$ (Fig. 8). This difference is controlled by the model parameters, in particular, by the association constant $K_a \sim e^{|\varepsilon|}$: $\Delta\rho$ grows when increasing the value of $K_a$ and decreases when increasing the Flory-Huggins parameter $\chi_2$ characterizing the incompatibility between the B polymer and the first solvent.

The values of $\Delta\rho$ are not high (up to 0.1) in the considered case of small fractions of bonded monomer units, and the dependence of $\Delta\rho$ on $\alpha$ is almost linear. Hence with increasing pH, the difference $\Delta\rho$ goes to zero. This effect originates mainly from the decrease in $\rho_{\text{right}}$. The presence of mobile ions creates the contribution to the osmotic pressure proportional to the concentration of the ionisable polymer. This contribution is very small in the external solution of the A polymer but higher in the gel which leads to an increase in the concentration of the associating B polymer (a decrease in $\rho_{\text{right}}$) at the right boundary. Assuming that the composition difference $\Delta\rho$ of the hydrogel plays the role of $\Delta\rho^*$ in the dynamic model (Eqs. (12), (13)), we predict a sharp change of the growth rate by fine changes in polymer ionization.

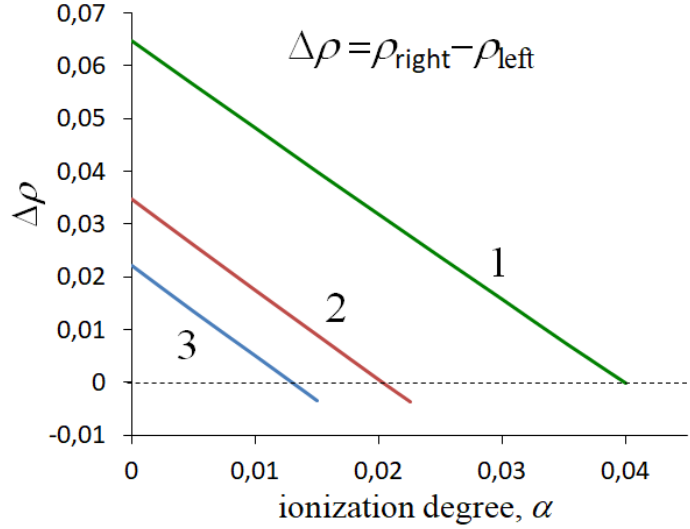

Figure 8. The dependence of the difference in the equilibrium fractions of the A polymer at the boundaries, $\Delta\rho = \rho_{right} - \rho_{left}$, on the ionization degree $\alpha$ for (1) $\tilde{K}_a = 3.5$, $\chi_2 = 0.65$; (2) $\tilde{K}_a = 3.3$, $\chi_2 = 0.65$, and (3) $\tilde{K}_a = 3.3$, $\chi_2 = 0.75$. The other parameters: $N_1 = 100$, $N_2 = 70$, $\varphi_{1out} = 0.01$, $\varphi_{2out} = 0.1$. The dashed horizontal line corresponds to the condition $\rho_{left} = \rho_{right}$.

It is worth noting that the polymer ionization also reduces the number of monomer units capable of bonding, as if the association constant $\tilde{K}_a$ were replaced by $\tilde{K}_a(1-\alpha)$ in the free energy expression (compare Eqs. (1) and (17)). This should lead to an additional decrease in growth rate of the membrane as predicted by Eq. (12). However, this effect is expected to be small for weakly ionized polymers.

The stationary regime of membrane growth with the exponential dependence $h \sim t^{1/2}$ is observed for the almost neutral polymers only. For the ionization degree around 10 %, the exponent $\beta$ in the time dependence is less than ½ (Fig. 6) revealing a possible non-stationary regime. If we assume that the polymer concentrations at the membrane boundaries keep their equilibrium values during the process, then the average polymer diffusivity should decrease with the membrane thickness. Such scale-dependent effects can indicate a loose and inhomogeneous structure of the ionized gel with reduced concentration of hydrogen bonds.

4. **Conclusion**

Membranes formed by associating polymers at liquid-liquid interfaces have distinctive properties relative to membranes formed during interfacial polymerization. In particular, they demonstrate self-healing ability and sensitivity to pH changes that control polymer ionization. We have observed experimentally that the membrane growth slows down significantly with increasing pH, whereas the ionization degree is quite small. The ionization degree of PMAA depending on pH in aqueous solutions has been determined in the range of interest.

For almost neutral polymers, the law of film growth resembles the stationary regime of diffusion-controlled process described earlier for interfacial polymerization. However, the process is principally different with the whole polymer chains as diffusive objects and the formation of reversible bonds everywhere in the film as a reaction. We have suggested a model of membrane formation in the process of polymer association, where the polymer absorption and the film growth are sustained by polymer interdiffusion giving rise to the same power-law $h \sim t^{1/2}$ in a stationary regime. The growth rate is determined by the association constant, polymer lengths and mobilities, and the equilibrium membrane compositions at the boundaries. These compositions can be calculated from the conditions of thermodynamic equilibrium between the boundary layers and the external solutions, where the membrane is considered as an aqueous solution of the polymer mixture.

Polymer ionization makes some PMMA monomer units unable to form H-bonds. Besides, it produces counterions in the hydrogel film and in the outer aqueous solution. A small fraction of ionized monomer units causes only small changes in the concentration of H-bonds and the equilibrium gel parameters. However, if the composition gradient is small, such changes can reduce this gradient to zero and thus terminate polymer interdiffusion.

The membrane growth was approximated by the dependence $h \sim t^{\beta}$, where the exponent $\beta$ decreased with increasing degree of ionization. The value $\beta = 0.5$ corresponds to a stationary interdiffusion-controlled process in the membrane of constant density. The values $\beta < 0.5$ reveal a non-stationary process, where interdiffusion and polymer absorption slow down over time (and thickness) more than in a stationary case. Such regime probably indicates that the membranes are inhomogeneous and they become more and more inhomogeneous with increasing ionization. A dynamical experimental or numerical study of the local structure of hydrogel films, in situ, would be a clear way to advance further our understanding of these objects, but remains an experimental challenge, as of today.

**AUTHOR INFORMATION**

**Corresponding Authors**

E-mail: elena.govorun@espci.fr

E-mail: cecile.monteux@espci.fr

**Notes**

There are no conflicts to declare.

**ACKNOWLEDGMENT**

The experimental studies were supported by the ANR JCJC INTERPOL (grant number ANR-12-JS08-00070). E. N. G. is thankful to the French National program Pause for the financial support.